\title{\normalfont A Superior but Equally Convenient Alternative to the Steady-State Approximation and Secular Equilibrium}
\author{
        {\em K. Razi Naqvi\/} \\
                Department of Physics,
        Norwegian University of Science and Technology\\
        NO-7491 Trondheim, Norway\\[3ex]}

\date{}

\documentclass[11pt,reqno]{article}
\usepackage{amsmath,amssymb,amsfonts} 
\usepackage{makeidx}  
\usepackage{graphicx} 
\usepackage{epsfig}
\usepackage[sort&compress,numbers]{natbib}
\usepackage{pstricks}
\usepackage{soul}  
\usepackage[dvips]{curves}  
\usepackage{xcolor}
\usepackage{listings}
\usepackage{mathtools}
\usepackage{mathrsfs}

\usepackage{multicol,caption}
\usepackage{float} 
\newenvironment{Figure}
  {\par\medskip\noindent\minipage{\linewidth}}
  {\endminipage\par\medskip}
\usepackage{pstricks, pst-text}                  
\usepackage{color}
\usepackage{fancyhdr}

\usepackage{mathpazo}
\usepackage{multicol,caption}
\usepackage{booktabs}
\makeatletter
\AtBeginDocument{
\renewcommand\NAT@open{\color{green}[}}
\usepackage{url}
\usepackage[bottom]{footmisc}
\def\d{{\,\rm d}}
\def\e{\hspace{0.05ex}{\rm e\hspace{0.05ex}}}

\def\d{{\,\rm d}}
\def\e{{\rm e}}

\begin{document}

\maketitle

\begin{center}
\section*{\fontfamily{phv}\selectfont\normalsize Abstract}
\label{sec:Abstract}
\end{center}
\vspace{1ex}

The steady-state approximation (hereafter abbreviated as SSA) consists in setting $\d y/\d t=0$, where $y$ denotes the concentration of a short-lived intermediate subject to first-order decay with a rate constant $k$. The sole reason for enforcing SSA is to convert the rate equation for $y$ into an algebraic equation. The conditions under which SSA becomes trustworthy are now well understood, but a firm grasp of the physical content of the approximation requires more maturity than few teachers, let alone their students, may be expected to possess; furthermore, there is no simple way to gauge the accuracy of the approximation. The purpose of this note is to demonstrate that a better, but equally simple, approximation results if, instead of setting $\d y/\d t$ to zero, one substitutes $y(t+\tau)$ for  $y+\tau \d y/\d t$, where $\tau=1/k$; SSA is a cruder approximation because it neglects the second term. For systems modelled as damped harmonic oscillators, the ``reverse Taylor approximation" can be extended by retaining one more term in the Taylor expansion. The utility of the approximation (or its extension) is demonstrated by examining the following systems: radioactive equilibria, Brownian motion, dynamic response of linear first- and second-order systems. \\[7ex]


\begin{multicols}{2}

\vspace*{-8ex}

\section{Introduction}

More than a hundred years have elapsed since Rutherford introduced the terms\linebreak ``secular equilibrium" and ``transient equilibrium"  \cite[p.~429]{Rutherford1913Radioactive}. The confusion surrounding these terms has long been a source of chagrin and concern to educators \cite{Prince1979JNuclMed}, and there appeared as late as 2004, a publication  bearing the title ``The concepts of transient and secular equilibrium are incorrectly described in most textbooks, and incorrectly taught to most students and residents"; there is still room, in the present author's opinion, for another article with a similar title but the words ``transient and secular equilibrium"  replaced by ``steady state approximation". The purpose of this article is to discuss these terms in the light of an approximation that surpasses SSA without requiring a deeper knowledge of calculus; apart from being transparent, this approach permits a reliable estimate of the error caused by its use, and it is the author's hope that this note will dispel the fog that has been preventing clear thinking about SSA and secular/transient equilibria.

\section{Statement of the new approximation}

We will denote time, the independent variable, by the symbol $t$, and will use, when convenient, the abbreviation ${\mathcal D}\equiv \d /\d t$. We will begin by considering differential equations (ODE's) involving only ${\mathcal D}Y$. The symbol $Y_{\rm{st}}$ will be used for the result obtained by using SSA.

Let $Y$ denote the quantity to which the new approximation is to be applied, and let its time rate of change be described by the ODE (in which $k$ is independent of $t$)
\begin{subequations}\label{eq:RateEq001}
\begin{align}
\frac{\d Y}{\d t} &= f(t)- kY,				\label{eq:RateEq001a}\\
\noalign{\noindent\mbox{which can be easily rearranged as}}
Y+\frac{1}{k}\frac{\d Y}{\d t} &=\frac{f(t)}{k}	\label{eq:RateEq001b}
\end{align}
\end{subequations}
Upon setting ${\mathcal D}Y=0$ in Eq.~(\ref{eq:RateEq001a}), one gets
\begin{equation}
Y_{\rm{st}}=\frac{f(t)}{k}=\tau f(t),\qquad (\tau=1/k).
\end{equation}

Recalling that $Y(t+\tau)$ can be expanded as a Taylor series
\begin{align}
Y(t+\tau) = Y(t)&+ \tau \frac{\d Y}{\d t}+ \nonumber\\
&\hspace{1ex} \frac{\tau^2}{2} \frac{\d^2 Y}{\d t^2}+\cdots ,
\end{align}
we go on to introduce two versions of what will be called the {\em reverse  Taylor approximants\/}:
\begin{subequations}
\begin{align}
Y^{(0)}(t+\tau)	&= Y(t)\\
Y^{(1)}(t+\tau) &= Y(t)+ \tau \frac{\d Y}{\d t}. 
\end{align}
\end{subequations}
The zero-order approximant, since it results from putting ${\mathcal D}Y=0$, coincides with $Y_{\rm{st}}$; in contrast, the first-order approximant retains ${\mathcal D}Y$, and replaces the left-hand side of Eq.~(\ref{eq:RateEq001b}) by $Y^{(1)}(t+\tau)$, which gives
\begin{subequations}
\begin{align}
Y^{(1)}(t+\tau)	&= \tau f(t).\\
\noalign{\noindent\mbox{Whence follows the result }}
Y^{(1)}(t)&=\tau f(t-\tau).
\end{align}
\end{subequations}

\subsection{Range of validity and accuracy}

The first-order reverse Taylor approximant (for short, RTA-1) does not contain $Y_0$, the initial value of $Y$, which may have any finite value ($Y_0\geq 0$). Since Eq.~(\ref{eq:RateEq001a}) is linear, we know that its solution must contain a term $Y_0\exp({-k t})$, and this term will not be negligible unless we confine attention to times which are longer than a few lifetimes of the labile species (and we will take {\em few\/} to mean {\em five\/}). We conclude therefore that TA-1 is inapplicable at early times ($t\leq 5 \tau$).

The accuracy of the approximation is determined by the fact that we have ignored terms containing ${\mathcal D}^nY$ ($n\geq 2$). If each term in the Taylor expansion is an order of magnitude smaller than its predecessor, we are entitled to expect an accuracy of around 1\%;  formally, we express this restriction as
\begin{equation}\label{eq:accuracy}
\mu^n {\mathcal D}^nY \leq \varepsilon \mu^{n-1}{\mathcal D}^{n-1}Y, \qquad (n\geq 1),
\end{equation}
where $\varepsilon \approx 10^{-1}$ and $D^0=1$.
The following sections will demonstrate how the above inequality can be applied to deal with specific cases.

\section{Application to successive disintegration}

The frist-order Taylor approximation will now be applied to the nuclear reaction
\begin{equation}\label{eq:ReacScheme}
A_1\xrightarrow[]{\quad \lambda_1\quad}A_2\xrightarrow[]{\quad \lambda_2\quad}A_3, \tag{Scheme~1}
\end{equation}
which has only one intermediate, and obeys the following set of rate equations \cite[p.~98]{Kaplan1955Nuclear}:
\begin{align}
\frac{\d N_1}{\d t}&= 	-\lambda_1 N_1,			\label{eq:RateEq01}\\
\frac{\d N_2}{\d t}&= 	\lambda_1N_1 -\lambda_2 N_2,	\label{eq:RateEq02}\\
\frac{\d N_3}{\d t}&=	\lambda_2 N_2,
\end{align}
where $N_i$ stands for the number of atoms of the radioactive species $A_i$ at time $t$; the initial condition will be taken as: $N_i=N_i^0$ at $t=0$. 

Since the right superscript will now be used for denoting the initial value of $N_i$, the order of the approximation will appear as the left superscript. It will avoid clutter if we use symbols for the disintegration constants and their reciprocals that do not have subscripts; accordingly, we introduce the notations
\begin{equation}
\lambda=\lambda_1,\quad \kappa=\lambda_2,\quad \mu=\kappa^{-1}
\end{equation}
and replace Eqs.~(\ref{eq:RateEq01}) and (\ref{eq:RateEq02}) by the pair shown below
\begin{align}
\frac{\d N_1}{\d t}&= 	-\lambda N_1,			\label{eq:RateEq07}\\
N_2+\mu\frac{\d N_2}{\d t}&= \frac{\lambda N_1}{\kappa} ,		\label{eq:RateEq08}
\end{align}
Replacing the left-hand side of Eq.~(\ref{eq:RateEq08}) by the first approximant $^{(1)\!}N_2 (t)$, we obtain
\begin{align}
^{(1)\!}N_2 (t+\mu)&= \frac{\lambda}{\kappa} N_1(t)\\
\therefore\quad  ^{(1)\!}N_2 (t)&= \frac{\lambda}{\kappa} N_1(t-\mu).\\
\noalign{\noindent\mbox{But, since  $N_1 (t)= N_1^0\,\e^{-\lambda t}$, we have}}
^{(1)\!}N_2 (t)&=\frac{\lambda}{\kappa} N_1^0\exp[-\lambda(t-\mu)].\label{eq:TA-I4N2}
\end{align}

As stated above, the solution should be used only when $t\geq 5 \mu$. It also follows from 
Eq.~(\ref{eq:TA-I4N2}) that 
\begin{equation}
\mu^2{\mathcal D}^2\left [{}^{(1)\!}N_2 (t)\right ]=  \lambda^2\mu^2\cdot{}^{(1)\!}N_2 (t)
\end{equation}
In view of the inequality (\ref{eq:accuracy}), we will impose the demand that terms involving second and higher powers of $\lambda/\kappa=\lambda\mu$ will be considered negligible in comparison with unity. Given the formula
\begin{equation}
1+\lambda\mu + (\lambda\mu)^2 + (\lambda\mu)^3 + \cdots =\frac{1}{1-\lambda \mu},
\end{equation}
one sees that it will be legitimate to use the approximations
\begin{equation}\label{eq:accuracy2}
\e^{\lambda \mu}\approx 1+\lambda \mu \approx \frac{1}{1-\lambda\mu},\qquad (\lambda\mu\ll 1),
\end{equation}
because $\e^{\lambda \mu}$, $1+\lambda \mu$ and $(1-\lambda \mu)^{-1}$ differ only by terms of order $\lambda^2\mu^2$.

We return now to Eq.~(\ref{eq:TA-I4N2}) and carry out the following manipulations
\begin{align}
^{(1)\!}N_2 (t)&= N_1^0\,\frac{\lambda}{\kappa}\,\e^{-\lambda(t-\mu)}=\e^{-\lambda t}\e^{\lambda\mu}	\nonumber\\
&\approx N_1^0\,\frac{\lambda}{\kappa}\,\e^{-\lambda t} \frac{1}{1-\lambda\mu}		\nonumber\\
&= N_1^0\,\frac{\lambda}{\kappa}\,\e^{-\lambda t} \frac{1}{1-({\lambda}/{\kappa})}	\nonumber\\
&=\frac{\lambda}{\kappa-\lambda} N_1^0\,  \e^{-\lambda t} .
\end{align}
The time has now come to revert to the original notation, to redress the above result as
\begin{equation}
N_2\approx N_2^{(1)}=
\frac{\lambda_1}{\lambda_2-\lambda_1} N_1^0\,  \e^{-\lambda_1 t}=
\frac{\lambda_1N_1}{\lambda_2-\lambda_1},
\end{equation}
and to recall the exact solution given in his Eq.~(10-19) by Kaplan \cite{Kaplan1955Nuclear}:
\begin{align}
N_2&=\frac{\lambda_1}{\lambda_2-\lambda_1}N_1^0\,(\e^{-\lambda_1 t}-\e^{-\lambda_2 t})+\nonumber\\
&\hspace{12ex} N_2^0\,\e^{-\lambda_1 t}. \tag{Kaplan: 10-19}
\end{align}
\vspace{2ex}
Kaplan writes: 

\leftskip=20pt
{\small
A \ldots state of affairs, called {\em transient equilibrium\/}, results if the parent is longer-lived than the daughter ($\lambda_1 <\lambda_2$), but the half-life of the parent is not very long. \ldots After $t$ becomes sufficiently large, $\e^{-\lambda_2 t}$ becomes negligible compared with $\e^{-\lambda_1 t}$, and the number of atoms of the daughter becomes
\begin{equation}
N_2\approx \frac{\lambda_1}{\lambda_2-\lambda_1}N_1^0\,\e^{-\lambda_1 t}. \tag{Kaplan: 10-35}
\end{equation}
Thus, the daughter eventually decays with the same half-life as the parent. Since $N_1^0\,\e^{-\lambda_1 t}=N_1$, it follows from Eq.~(10-35) that
\begin{equation}
\frac{N_1}{N_2} =\frac{\lambda_2-\lambda_1}{\lambda_1}. \tag{Kaplan: 10-36}
\end{equation}
}

\leftskip=0pt

If we envisage a system where $\lambda_2\gg \lambda_1$, with $1/\lambda_1$ so very large that the decay of the species $A_1$ during a much smaller time interval (which may be the duration of the experiment or the life span of the observer) is truly negligible, a state of affairs, called {\em secular equilibrium\/}, is reached, characterized by the relation
\begin{equation}
\lambda_1N_1=\lambda_2N_2,
\end{equation}
which can be obtained directly from Eq.~(\ref{eq:RateEq02}) by invoking SSA and setting
\begin{equation}
\frac{\d N_2}{\d t}=0
\end{equation}

Though the ratio $N_1/N_2$ is found to be a constant whether one uses the steady-state approximation (equivalent to secular equilibrium) or RTA-1 (equivalent to transient equilibrium), the value of the ratio is not the same, being $\lambda_2/\lambda_1$ in the former case and $(\lambda_2 - \lambda_1)/\lambda_1$ in the latter case.

The reader will not have failed to notice that RTA-1 led us to the desired result {\em without actually solving the rate equation\/} [namely, Eq.~(\ref{eq:RateEq02})], and that this result is more accurate than that furnished by SSA, which corresponds to RTA-0 (zero-order reverse Taylor approximation).

\section{Dynamic response of a first-order sensor}

The dynamic response of a first-order sensor is modelled by the ODE
\begin{equation}
\frac{\d Y}{\d t}+ kY=f(t),
\end{equation}
were $Y$ is the output of the sensor and $f(t)$ the input. The constant $\tau=1/k$ is called the {\em time constant\/}, but the term {\em response time\/} will be equally apt in the present context.

We will now investigate the response (output) of a first-order sensor to a ramp input:  $f(t)=0$ for $t<0$, and $f(t)=At$ for $t\geq 0$, where $A$ is a constant. The initial condition is $Y=0$ at $t=0$

The expression for TA-1 can be stated immediately:
\begin{equation}
Y^{(1)}(t)=
\frac{A}{k}\left [t-\frac{1}{k}\right ]=
\frac{A}{k^2}(kt-1)
\end{equation}
\begin{Figure} 
  \centering
\includegraphics[width=\textwidth]{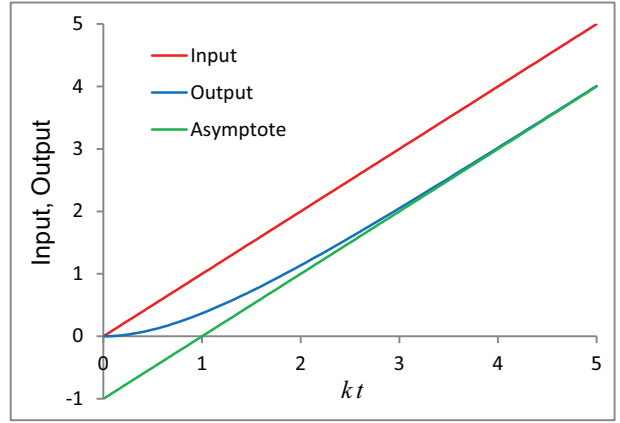}
  \captionof{figure}{The response of a first-order sensor to a ramp input. For other details, see the text.}
  \label{Figure:FigureOne}
\end{Figure}

The exact solution is easily found to be
\begin{align}
Y(t)&=\frac{A}{k}\left [t-\frac{1}{k}(1-\e^{-kt})\right ], \label{eq:Ramp02}
\end{align}
and it will be instructive to split it into an asymptotic part ($Y_{\rm as}$) and a transient part ($Y_{\rm tr}$) that becomes negligible when $kt>5$:
\begin{align}
Y(t)&=\underbracket[0.5pt]{\frac{A}{k^2}(kt-1)}_{\displaystyle{=Y_{\rm as}}} + \underbracket[0.5pt]{\frac{A}{k^2}\e^{-kt}}_{\displaystyle{=Y_{\rm tr}}}
\end{align}

The steady state approximation, which consists in setting ${\mathcal D}Y=0$, leads to the result
\begin{equation}\label{eq:RampSSA}
Y \approx Y_{\rm st}=\frac{At}{k}=Y^{(0)},
\end{equation}
is always inferior to $Y^{(1)}$, except when $kt< 1$, but at such short times neither of these approximations should be used.

Figure~\ref{Figure:FigureOne} shows plots (with $A/k^2=1$) of the ramp input, output $Y$ and its asymptotic part $Y_{\rm as}$.   

The first-order reverse Taylor approximation ignores terms containing ${\mathcal D}^nY$ (for $n\geq 2$), and application of ${\mathcal D}^2$ to Eq.~(\ref{eq:Ramp02}) shows that
\begin{equation}\label{eq:DynamicRampD2y}
{\mathcal D}^2Y =A\e^{-kt},
\end{equation}
from which one may conclude that the higher derivatives do indeed die out when $kt\geq 5$. For a ramp input, RTA-1 (that is, $Y^{(1)}$) coincides with the asymptotic form of the solution, but this is not a general result. To demonstrate this, it will be enough to consider a parabolic input $f=Bt^2$. The expression for RTA-1 now comes out to be
\begin{align}
Y^{(1)}&=\frac{B}{k}\left (t-\tau\right )^2&&(t\geq 5\tau)\\
&=\frac{B}{k^3}\left [k^2t^2-2kt+1\right ]&&(kt\geq 5)
\end{align} 
Though the asymptotic part of the exact solution, show below,
\begin{equation}\label{eq:OutParab01}
Y=\underbracket[0.5pt]{\frac{B}{k^3}\left (k^2t^2-2kt+2\right )}_{\mbox{\footnotesize asymptotic part}}-\frac{2B}{k^3}\e^{-kt}
\end{equation}
is not identical with $Y^{(1)}$, the difference occurs at the expected level of accuracy, namely the constant term, which becomes unimportant in the time range of interest, $kt >5$.

\section{Brownian motion}

For one-dimensional brownian motion in a field-free infinite region, Einstein was able to show that the diffusion equation
\begin{equation}\label{eq:Einstein-DE}
\frac{\partial F}{\partial t}=D\frac{\partial^2 F}{\partial x^2},
\end{equation}
in which $F(x,t)$ is the probability density and $D$ the diffusion coefficient, implies the following result for the mean-squared displacement of the diffusing particle:
$$
\overline{x^2}\equiv \int_{-\infty}^{\infty} x^2 F(x,t)\d x=2Dt.
$$

Ornstein \cite{Ornstein1919KNAW} appears to have been the first to show that, when the inertia of the particle is taken into account, the mean-squared displacement is given by the formula
\begin{equation}\label{eq:OvW-x2}
\overline{x^2} = \frac{2D}{\beta} (\beta t-1+\e^{-\beta t}),
\end{equation}
where $\beta$, known as the velocity relaxation time, may be defined through the
the equation of motion
$$
\frac{\d v}{\d t} = -\beta t +  \mathscr{A}
$$
that is supposed to govern the time dependence of the velocity $v$ of a brownian particle subject to a random acceleration $\mathscr{A}$. Ornstein commented: ``As long as $\beta t$ is large in relation to $1-\e^{-\beta t}$ the formula of Einstein is thus the right one." From what has been said above, it becomes clear that the formula
$$
\overline{x^2}=2D(t-\beta^{-1})
$$
provides a much better description at $t\geq 5 \beta^{-1}$.  For more details, the reader is referred to some earlier works \cite{Unruh1980AJP,KRN1982PRL,KRN2005arXivLangevin}.

\section{Dynamic response of a second-order sensor}

The heart of a second-order sensor is a damped harmonic oscillator governed by the equation of motion shown below:
\begin{equation}\label{eq:EoM-Order2}
\left (\frac{1}{\omega^{2}}{\mathcal D}^2 + \frac{2\gamma}{\omega^{2}}{\mathcal D} + 1 \right )Y=\frac{1}{\omega^{2}}f(t).
\end{equation}
The key parameter in this context is the damping ratio $\zeta=\gamma/\omega$. The oscillator is said to be overdamped, critically damped, or underdamped according as $\zeta$ is larger than, equal to, or smaller than unity. In sensing applications, a frequency analysis of the sensor response shows that the optimum value of the damping ratio is $\zeta=1/\sqrt{2}$ \cite[p.~70]{Tong1960Theory}. It will now be shown that this is also the optimum value for providing the best Taylor approximant for the output

Our aim in this section is to replace the left-hand side of Eq.~(\ref{eq:EoM-Order2}) by $Y^{(2)}(t+\eta)$ (or RTA-2), which is defined below
\begin{equation}
Y^{(2)}(t+\eta)\equiv Y(t)+\eta \frac{\d Y}{\d t} + \frac{\eta^2}{2}\frac{\d^2 Y}{\d t^2} ,
\end{equation}
where 
\begin{equation}
\eta=\frac{2\gamma}{\omega^2}.
\end{equation}
We can now rephrase Eq.~(\ref{eq:EoM-Order2}) as
\begin{equation}
\left (1+\eta {\mathcal D}+\frac{1}{2\zeta^2}\frac{\eta^2}{2}
{\mathcal D}^2\right )Y=\frac{1}{\omega^2}f(t),
\end{equation}
and, if $2\zeta^2=1$, as
\begin{equation}
Y^{(2)}(t+\eta)= \frac{f(t)}{\omega^2}, 
\end{equation}
from which follows the result
\begin{equation}
Y(t)\approx Y^{(2)}(t)= \frac{f(t-\eta)}{\omega^2}.
\end{equation}
The task of ascertaining the accuracy of this result is left to the reader.

\section{Concluding remarks}

It has been shown above that, by reversing the direction in which Taylor's expansion is used traditionally, one can approximate an equation of the form ${\mathcal D}y+ky=f(t)$ as $y(t+k^{-1}y)=k^{-1}f(t)$, which is a substantial improvement, both in terms of accuracy and ease of interpretation, over the equation $y=k^{-1}f(t)$, obtained by the application of SSA to the same equation. The foregoing analysis also throws new light on the exact nature of SSA. For a clear statemet of the traditional explanation, I simply paraphrase Briggs and Haldane \cite{Briggs1925BiochemJ} by adapting their justification for SSA to \ref{eq:ReacScheme}: ``Since $N_2$ is always negligible compared with $N_1$ and $N_3$, its rate of change must, except during the first instant of the reaction, be negligible compared with theirs." According to the formulation developed above [see Eq.~(\ref{eq:RateEq08})], and presented briefly in a recent article \cite{Arellano2017IJCK}, SSA ignores (in comparison with $N_2$) the product $\mu(\d N_2/\d t)$, which is the change in $N_2$ during an interval of time $\mu$ (that is exceedingly short compared with the lifetime of the parent). The main point of this article is to demonstrate that there is, in fact, no need to ignore this term, since the reverse Taylor approximation allows one to accommodate it easily. Whence follows the conclusion that the time has come to regard SSA as a relic of the past, and start using the reverse Taylor approximation proposed above.


\begin{thebibliography}{10}

\bibitem{Rutherford1913Radioactive}
E.~Rutherford.
\newblock {\em Radioactive Substances and Their Radiations}.
\newblock Cambridge University Press, 1913.

\bibitem{Prince1979JNuclMed}
J.~R. Prince.
\newblock Comments on equilibrium, transient equilibrium, and secular
  equilibrium in serial radioactive decay.
\newblock {\em J. Nucl. Med.}, 20(2):162--164, 1979.

\bibitem{Kaplan1955Nuclear}
I.~Kaplan.
\newblock {\em Nuclear Physics}.
\newblock Addison-Wesley, Reading, Massachusetts, 1955.

\bibitem{Ornstein1919KNAW}
L.~S. Ornstein.
\newblock On the {B}rownian motion.
\newblock {\em Proc. Acad. Amst.}, 21(1/2):97--108, 1919 (original published in
  1917).

\bibitem{Unruh1980AJP}
Henry Unruh.
\newblock Experimental study of {B}rownian motion in the limit of small time
  intervals by means of an ideal gas simulator.
\newblock {\em Am. J. Phys.}, 48(10):818--820, 1980.

\bibitem{KRN1982PRL}
K.~Razi Naqvi, K.~J. Mork, and S.~Waldenstr\o{}m.
\newblock Reduction of the fokker-planck equation with an absorbing or
  reflecting boundary to the diffusion equation and the radiation boundary
  condition.
\newblock {\em Phys. Rev. Lett.}, 49:304--307, 1982.

\bibitem{KRN2005arXivLangevin}
K.~Razi Naqvi.
\newblock The origin of the {L}angevin equation and the calculation of the mean
  squared displacement: Let's set the record straight.
\newblock arXiv:physics/0502141v1 [physics.chem-ph], (2005).

\bibitem{Tong1960Theory}
K.~N. Tong.
\newblock {\em Theory of Mechanical Vibration}.
\newblock Wiley, New York, 1960.

\bibitem{Briggs1925BiochemJ}
G.~E. Briggs and J.~B.~S. Haldane.
\newblock A note on the kinetics of enzyme action.
\newblock {\em Biochem. J.}, 19(2):338--339, 1925.

\bibitem{Arellano2017IJCK}
Juan~B. Arellano, Elena Mellado-Ortega, and K.~Razi Naqvi.
\newblock The {ORAC} assay: Mathematical analysis of the rate equations and
  some practical considerations.
\newblock {\em Int. J. Chem. Kin.}, 49(6):409--418, 2017 (Sup-3).

\end{thebibliography}
%

\end{multicols}

\end{document}